\documentclass[prb,aps,showpacs,floatfix,superscriptaddress,groupedaddress,amsmath,twocolumn]{revtex4}

\usepackage[dvips]{graphicx}
\usepackage{dcolumn}
\usepackage{bm}
\usepackage{epsfig}
\usepackage{amssymb}

\newcommand{\be}{\begin{equation}}
\newcommand{\ee}{\end{equation}}
\newcommand{\bea}{\begin{eqnarray}}
\newcommand{\eea}{\end{eqnarray}}
\newcommand{\bean}{\begin{eqnarray*}}
\newcommand{\eean}{\end{eqnarray*}}

\begin{document}

\title{ Ab-initio calculations for the $\beta$-tin diamond transition in Silicon: comparing theories with experiments}

\author{Sandro Sorella}
\affiliation{International School for Advanced Studies (SISSA), Via Bonomea 265, 34136 Trieste, Italy}
\affiliation{ DEMOCRITOS Simulation Center
CNR-IOM Istituto Officina dei Materiali, 34151, Trieste, Italy}
\author{Michele Casula}
\affiliation{CNRS and Institut de Min\'eralogie et de Physique des Milieux condens\'es,
case 115, 4 place Jussieu, 75252, Paris cedex 05, France}
\author{Leonardo Spanu}
\affiliation{Department of Chemistry, University of California at Davis, Davis, CA, USA}
\author{ Andrea Dal Corso}
\affiliation{International School for Advanced Studies (SISSA), Via Bonomea 265, 34136 Trieste, Italy}
\affiliation{ DEMOCRITOS Simulation Center
CNR-IOM Istituto Officina dei Materiali, 34151, Trieste, Italy}

\date{\today}
\begin{abstract}
We investigate the pressure-induced metal-insulator transition from
diamond to $\beta$-tin in bulk Silicon, using quantum
Monte Carlo (QMC) and density functional theory (DFT) approaches. We
show that it is possible to efficiently describe many-body effects,
using a variational wave function with  an
optimized Jastrow factor and a Slater determinant. Variational results
are obtained with a small computational cost and are further improved
by performing diffusion Monte Carlo calculations and an explicit
optimization of molecular orbitals in the determinant. Finite
temperature corrections and zero point motion effects are included by
calculating phonon dispersions in both phases 
at the DFT level. 
Our results indicate that the theoretical QMC (DFT) 
transition pressure is significantly
larger (smaller) than the accepted experimental value. We discuss the limitation
of DFT approaches due to the choice of the exchange and correlation
functionals and the difficulty to determine consistent 
pseudopotentials within the QMC framework,  
a limitation that may significantly affect the accuracy of the technique. 
 
\end{abstract}

\maketitle
\section{Introduction}
\label{intro}

The prediction of material behavior under pressure is of great relevance in several branches of science, 
ranging form material science to planetary physics\cite{emu}.  The experimental determination of pressure effects 
on the phase diagram can be rather complicated even for simple inorganic crystal, 
due to the related extreme conditions or the existence of subtle phenomena, such as hysteresis. 
In this respect, {\it ab-initio} calculations are complementary methods for determining phase diagrams 
and understanding material phases in a large range of pressures and temperatures.
 
The density functional theory (DFT)
has been widely used 
for describing material behavior under pressure. 
However, 
when a transition is accompanied by a drastic change in the electronic structure,
non-canceling errors in the two phases can lead to a significant bias in the predicted transition pressure. 
That is the case of bulk Silicon (Si), where the
diamond-to-$\beta$-tin transition is associated with
a semiconductor-to-semimetal electronic change.
For this reason, the Si diamond-to-$\beta$-tin transition 
has been used for testing and benchmarking new {\it ab-initio} numerical approaches for extended systems.
Its first order nature makes it a difficult case for
the experiments\cite{exp1,exp3,exp4}.
The accepted experimental values for the transition pressure are in
between $10$ and $12.5$ GPa at room temperature,
the difference could be ascribed to
non hydrostatic conditions\cite{exp2}, non quenched metastable
phases\cite{exp5}, and
presence of lattice defects\cite{exp6}.

Theoretical calculations based on DFT strongly depend on the choice of exchange and correlation (XC) functional. 
Calculations based on the local density approximation (LDA)
predict a zero temperature transition pressure $p_t$
in the range of $7.2$\cite{dalcorso}-$8.2$\cite{leemartin} GPa 
(this difference could be explained by the type of pseudopotential (PP)
used). The generalized gradient approximation (GGA) leads to $p_t$ in the
range of $12.2$\cite{moll}-$13.5$\cite{dalcorso} GPa
with the Perdew-Wang functional\cite{pw91}. 
Moreover, in the case of the PBE functional\cite{pbe}, a value of
$10.2$ GPa is obtained in Ref.~\onlinecite{leemartin}.
Since the PBE functional fulfills a number of exact DFT properties,
this should be considered the state-of-the-art among the 
most accurate ab-initio functionals.
A careful investigation of the effect of the XC functional on the
transition pressure was done recently in Ref.~\onlinecite{cyrus},
where the authors show that the inclusion of non-local exchange 
in the XC functional leads to a significant improvement of the
estimate of the transition pressure. 
All the calculations are performed
at zero temperature and therefore a comparison with experiments is
possible only after including zero-point 
motion and finite temperature effects. This accounts 
for a significant reduction of
the transition pressure,  as estimated 
in Ref.\onlinecite{temp_dep} 
by means of the LDA functional, with a correction larger than 1Gpa. 
Moreover the explicit inclusion of non linear core corrections (NLCC)
in pseudopotential calculations further reduces the transition pressure.

Quantum Monte Carlo (QMC) methods can be an alternative to DFT-based approaches. 
In the past years, many authors have shown practical applications 
of QMC methods for computing the energetics of extended systems\cite{review,graphite}, 
and predicting crystal phases under pressure\cite{ferro,alfe}. In a early work, Alf\'e {\it et al.}\cite{alfe} 
used diffusion Monte Carlo (DMC) for investigating the Si diamond-to-$\beta$-tin transition. 
They calculated a QMC transition pressure of $16.5$ GPa,
namely  $4$-$5$ GPa larger than  the experimental range. 
The discrepancy was attributed to the fixed-node (FN) approximation\cite{fn}, 
since the other source of errors (time step error, pseudopotential locality error, size effects) were considered negligible.  
More recently, Purwanto {\it et al.}\cite{purwanto} 
obtained a transition pressure
of $12.6$ using the auxiliary-field QMC (AFQMC) method with the phaseless
approximation to cure the sign problem, whose bias has been shown
to be very small\cite{shiwei_afqmc}.
The very recent DMC calculation in Ref.~\onlinecite{cyrus},
carried out with the most advanced 
optimization\cite{sandro_opt} and size-extrapolation\cite{KZK} methods,
gives a transition pressure of $14$ GPa in substantial agreement with
the AFQMC one, although the latest DMC value is affected by a larger
uncertainty  ($1$GPa).

In the present work, we address the problem of the Si diamond-to-$\beta$-tin zero temperature transition by
performing variational Monte Carlo (VMC) calculations for the total energy of the two crystal structures. 
VMC is not affected by the sign problem and has proven 
to be a reliable approach for several systems\cite{review,beaudet,casula_mol,graphite}. 
The accuracy of the method depends entirely on the choice of the variational wave function, and the capability of finding
its optimal form. In this work, we show that with a relative simple parameterization of the wave function (WF) we are able to describe 
correlation effects across the metal-insulator transition. 
Our wave function is a product of a Jastrow factor and a Slater determinant. 
The variational optimization of the Jastrow factor has a relative small computational cost\cite{sandro_opt}, 
and it reveals an accurate way to build-up correlation effects starting from an LDA calculation. 
 
The paper is organized as follows. In Sec.~\ref{comp_details} 
we review the properties of our variational wave function, 
we present a systematic study of the basis set used for the calculations, and we discuss
the correction of the finite size errors.
In Sec.~\ref{corrections} we review 
sources of errors not directly estimated at the QMC level, such as 
the quantum and thermal lattice energies, and the 
accuracy of the PP's. This will help us 
to make a fair comparison between our results and the most significant
experimental and theoretical findings reported in literature.
In Sec.~\ref{results} we report our results, while 
in Sec.~\ref{conclusions} we draw our conclusions.  

\section{Computational Details}
\label{comp_details}

In our investigation we have carried out 
DFT calculations with the
Quantum Espresso (QE)\cite{pwscf}, Wien2K\cite{wien2k} and {\it Qbox}\cite{qbox} packages,
 and we used the {\it TurboRVB} code\cite{turboRVB} to perform
variational Monte Carlo (VMC) and 
Lattice Regularized Diffusion Monte Carlo (LRDMC)\cite{lrdmc} calculations.
In both DFT and QMC the system is described by an effective Hamiltonian
in the Born-Oppenheimer approximation (without quantum effects of
lattice vibrations)  with the core electrons  treated 
neon-core PP. 

For extensive volume (pressure)
dependent calculations, we used the relativistic Hartree-Fock PP's generated by
Trail and Needs\cite{trail_fe}, employed also in previous 
QMC calculations of Refs.~(\onlinecite{cyrus,alfe}).
To check the impact of the PP approximation we carried out 
single-volume calculations using the Burkatzki-Filippi-Dolg energy-adjusted PP's\cite{bfd} and the ones generated
by the atomic PWSCF code\cite{pwscf} with the Troullier-Martins
construction\cite{troullier_martins} from relativistic LDA and PBE
atomic calculations. 
The PP's are expanded in a semilocal form in terms of
Gaussians\footnote{The pseudopotentials not available in the semilocal
  Gaussian form, as those generated by the atomic PWSCF code, have
  been fitted by a series of Gaussian functions. 
The discrepancy in the transition pressure
  obtained by the fitted pseudopotentials is less than 0.15 GPa from
  the original one.}.

\subsection{ Wave function and localized basis set}
\label{wave_function}

Our many-body wave function is the product of a Slater determinant and
a many-body Jastrow factor. Alternatively, in the case of open shell
systems, 
we use the Jastrow-correlated  Antisymmetrized Geminal Power (JAGP)\cite{casula_mol}.

The electron correlation is included in our wave function
through the Jastrow factor      
$J (\textbf{r}_1, \cdots , \textbf{r}_N) = \prod\limits_{i<j} \exp ( f( \textbf{r}_i, \textbf{r}_j ))$, 
where $f( \textbf{r},\textbf{r}^\prime)$ is assumed to depend only upon
two-electron coordinates, and $N$ is the total number of electrons.    
The function $f$ is expanded in a basis of Gaussian atomic orbitals $\bar \phi_i$, such that it reads:
\begin{equation}
\label{jas}
f( \textbf{r}, \textbf{r}^\prime) = \sum_{i,j}  g_{ij} \bar \phi_i (\textbf{r}) \bar \phi_j( \textbf{r}^\prime).
\end{equation}  
The convergence of this expansion is improved by adding an homogeneous term and a one-body contribution, 
thus satisfying the electron-electron and the electron-ion cusp conditions, respectively \cite{casula_atom,casula_mol}. 
The basis set used for the Jastrow includes $2s2p$ Gaussian orbitals.

For system with large number of electrons, we improve the efficiency of 
the optimization procedure and its computational cost,  adopting an
explicit parameterization of the Jastrow factor.  
Given two generic orbitals $ \bar \phi_i $ and $\bar \phi_j$, 
the variational coefficient $g_{ij}=g_{ij}(\textbf{R}_{ij})$ in Eq.~\ref{jas} depends only on the orbital symmetry  (in this case either s or p) and the distance vector $\textbf{R}_{ij}$ between the corresponding atomic centers.
 $g_{ij}$ is optimized without constraint when the two orbitals are localized on the same atom (i.e. $\textbf{R}_{ij}=0$).
Otherwise $g_{ij}(\textbf{R}_{ij})$ is parameterized in a way to
recover an isotropic large distance correlation.
For the sake of clarity, we define $\textbf{p}_j$ as the vector
containing the three $x,y,z$ p-orbital components centered at a given atomic position $\textbf{R}_j$, 
and with $s_j$ we indicate the s-wave orbital located at the same position.
By this notation we can write four possible isotropic invariant contributions 
for $ \textbf{R}_{ij} \ne  0$, such that:
\begin{eqnarray}
\label{parametrization}
\mathop{\sum_{l=s_i,\textbf{p}_i}}_{m=s_j,\textbf{p}_j} g_{lm}  \bar \phi_l \bar \phi_m
&=& f_1 ( R_{ij} )  \bar \phi_{s_i} \bar \phi_{s_j} 
\nonumber  \\  
&+& f_2(R_{ij}) ( \bar \phi_{s_i} \textbf{r}_{ij} \cdot
\bar \phi_{\textbf{p}_j}+  \textbf{p} \leftrightarrow s )  
\nonumber \\ 
&+& f_3(R_{ij})  ( \textbf{r}_{ij} \cdot  \bar \phi_{\textbf{p}_i} ) 
( \textbf{r}_{ij} \cdot  \bar \phi_{\textbf{p}_j} ) \nonumber  \\
&+& f_4(R_{ij})  ( \bar \phi_{\textbf{p}_i}  \cdot  \bar \phi_{\textbf{p}_j} ) \nonumber  \\
\end{eqnarray} 
where $R_{ij} = |\textbf{R}_{ij}|$, and
$\textbf{r}_{ij}=\textbf{R}_{ij}/R_{ij}$ is the unit vector connecting 
the atomic centers $i$ and $j$. The functions
$f_{p}(R_{ij})$ are polynomials which read:
\begin{eqnarray}
\label{scalar_functions}
 f_{p}(R_{ij}) & = &  C_0^p \log(R_{ij}) + \sum\limits_{k=1}^{k=2} C_k^p R_{ij}^{-k}  
~~~{\rm for}  ~ p=1,2 \nonumber \\
 f_{p}(R_{ij}) & = &   \sum\limits_{k=1}^{k=3} C_k^p R_{ij}^{-k}  
~~~{\rm for}  ~ p=3,4
\end{eqnarray}
The scalar functions in Eq.~\ref{scalar_functions} depend only on $12$
variational parameters, optimized via energy minimization\cite{srh}.

We verified the validity of the chosen parameterization (at long and
short distances), by a direct comparison with the case of a fully
optimized Jastrow factor (i.e. without parameterization).  
The expression in Eq.~\ref{parametrization} 
can be appropriate for physical long distance behaviors of the Jastrow factor, 
including the one recently speculated for describing the Mott insulator, 
and containing a long range $\log( R_{ij})$ term\cite{capello}.

The Slater determinant is obtained with $N/2$  molecular orbitals $\psi_j( \textbf{r})$, each doubly occupied 
by opposite spin electrons.
The orbitals $\psi_j(\textbf{r})$ are expanded in a Gaussian single-particle basis set $\{\phi _{i}\}$, 
centered on atomic nuclei, i.e. $\psi_j (\textbf{r})=\sum_{i}\lambda _{ji}\phi _{i}(\textbf{r})$.
The Slater determinant is build from LDA orbitals. LDA calculations are performed with a periodic Gaussian basis set 
(see Appendix \ref{gaussian_periodic} for its definition) by using the DFT code included in the {\it TurboRVB} package\cite{turboRVB}. 
This allows us to perform an efficient DFT calculation in exactly 
the same basis used in QMC and without employing the so called Kleinman-Bylander approximation on the PP's.

We carefully tested the effects of the basis set extrapolation 
on the total energy for both diamond and $\beta$-tin geometries. 
Following the systematically convergent method for accurate total energy 
calculations recently introduced in Ref.(\onlinecite{azadi}),  we have used a tempered basis set, where the Gaussian exponents $Z_i$ are defined as $ Z_i = \alpha \beta^{i}$ 
for $i=0,\ldots,n-1$ with $\alpha=Z_\textrm{min}$ and $\alpha \beta^{n} = Z_\textrm{max}$.
The parameters $n$, $Z_\textrm{min}$ and $Z_\textrm{max}$ are free. 
We verified that our basis set parameterization guarantees the same accuracy 
in both metallic and insulating phases for all investigated pressures.

For a given angular momentum $l$ ($\le 4$), 
we fixed the maximum number of Gaussians
according to the formula $ n_l = n_0 -2l $, inspired by the correlated
consistent basis set
approach\cite{correlation_consistent,correlation_consistent2}. It
follows that the maximum number $n$ of exponents is  
used only on the s-wave channel, where $n_0=n$. We studied the basis set
extrapolation with respect to $n$ by fixing $Z_\textrm{min}=0.05$ and $Z_\textrm{max}=10$. 

\begin{figure}[!ht]
\centering
\includegraphics[width=1.\columnwidth]{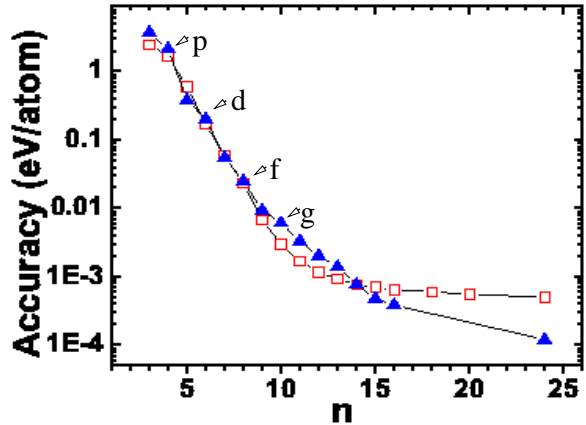}
\caption{
\label{basis}
\small{Accuracy of the DFT energy as a function of the 
basis set extension $n(=n_0)$ defined in the text. The number of Gaussians
in each angular channel $l$ is given by $n_l=n_0-2l$, as explained in
the text. In the plot, we
indicate with an arrow the value of $n$ when the corresponding angular channel has been switched on
in the basis set expansion. 
The red open squares  
 refer to the metallic $\beta$-tin phase at volume $13.081$\AA$^3/Si$ 
 and $c/a=0.54$, 
while the blue triangles to the diamond insulating phase at volume 
$20.036$\AA$^3/Si$.    
  } } 
\end{figure}

In Fig.~\ref{basis} we show the convergence of the DFT energy as a
function of $n (=n_0)$ for a system of $8$ Si atoms. 
We report the accuracy at $13.081$\AA$^3/Si$ and $c/a=0.54$ for the 
$\beta$-tin phase, and at volume $20.036$\AA$^3/Si$ for the diamond phase.
The accuracy in the energy is estimated by using a fully converged 
 reference energy of a plane-wave calculation with 100 Ry kinetic energy 
cutoff. This was obtained with the {\it Qbox} package\cite{qbox},  by using  
the same  PP's in the semilocal form as the ones used in our QMC calculations. 
An accuracy of $0.01$ eV/Si is sufficient 
to determine the equation of state (EOS) with an error well below the
experimental uncertainty. From this analysis it turns out that 
$n \ge 7$ and the inclusion of $d$-orbitals in the basis set 
guarantee an accuracy of $0.007$ eV/Si on the energy
difference between the two phases.

In the following, VMC and LRDMC production runs are performed using a basis set 
with $n_0=12$, $n_1=6$, and $n_2=4$.
The basis set exponents have been optimized at the DFT level. We found
the optimal parameters $Z_\textrm{min}=0.05$ and $Z_{max}=3.25$,
that minimize the DFT energy in both the diamond and $\beta-$tin
phases. 

All Jastrow parameters, including exponents, are obtained by  
means of  VMC energy minimization\cite{srh}.
Possible correlation 
effects not included by our Jastrow parameterization are recovered by performing LRDMC calculations. 
As a projector method, LRDMC allows one to obtain the best variational wave function 
with the same nodal structure as the initial variational wave
function, the so called  fixed node approximation (FN),
giving an upper bound of the true ground state energy even with non
local pseudopotentials\cite{lrdmc,lrdmc2}.

\subsection{Finite-size errors}
\label{finite_size}

Contrary to standard
DFT methods, QMC calculations have to be performed on a
supercell. Therefore, finite size (FS) effects can be a relevant
source of error in QMC calculations. Several methods have been
proposed for correcting FS errors. One source of FS errors arises from
the kinetic and Hartree term and can be treated by standard DFT
approach with $k-$point sampling. This is a genuine one-body
contribution, and can be corrected by $E^\textrm{DFT} - E^\textrm{DFT}_N$,
i.e. the difference between the DFT energy per atom with a fully converged
${\bf k}-$point mesh and the energy per atom of the supercell with finite
volume and number of electrons $N$.
 
The other source of errors (two-body terms) is related to the
finite size effects of  the exchange and correlation (XC) functionals, 
not explicitly included in $E^\textrm{DFT}_N$. 
We calculate the two-body term
corrections using the functional proposed by Kwee, Zhang and Krakauer
(KZK) \cite{KZK}. In Ref.~\onlinecite{KZK} the authors proposed to estimate
this type of FS error within the LDA framework, where the exchange and
correlation energy functional is replaced by the LDA functional
parameterized for a finite system, which keeps an explicit dependence
on the number of particles. Therefore, the total one- and two-body
correction is given by $\Delta^\textrm{KZK}= E^\textrm{DFT} - E^\textrm{DFT}_{N,\textrm{KZK}}$, where
$E^\textrm{DFT}_{N,\textrm{KZK}}$ is the DFT energy computed with the KZK functional
for $N$ electrons.

We observe that FS errors can be particularly relevant for open shell metallic systems. The AGP wave function approach allows to include many determinants in a effective way, removing the degeneracy of the open shell by  an appropriate fractional occupation of the degenerate levels. 
Within LDA, by using 
a negligible smearing in the occupation of the KS energy levels, 
the degenerate orbitals containing the HOMO are partially 
occupied with the same charge, and can be used consistently in the AGP wave function.

\begin{figure}
\centering
\includegraphics[width=1.0\columnwidth]{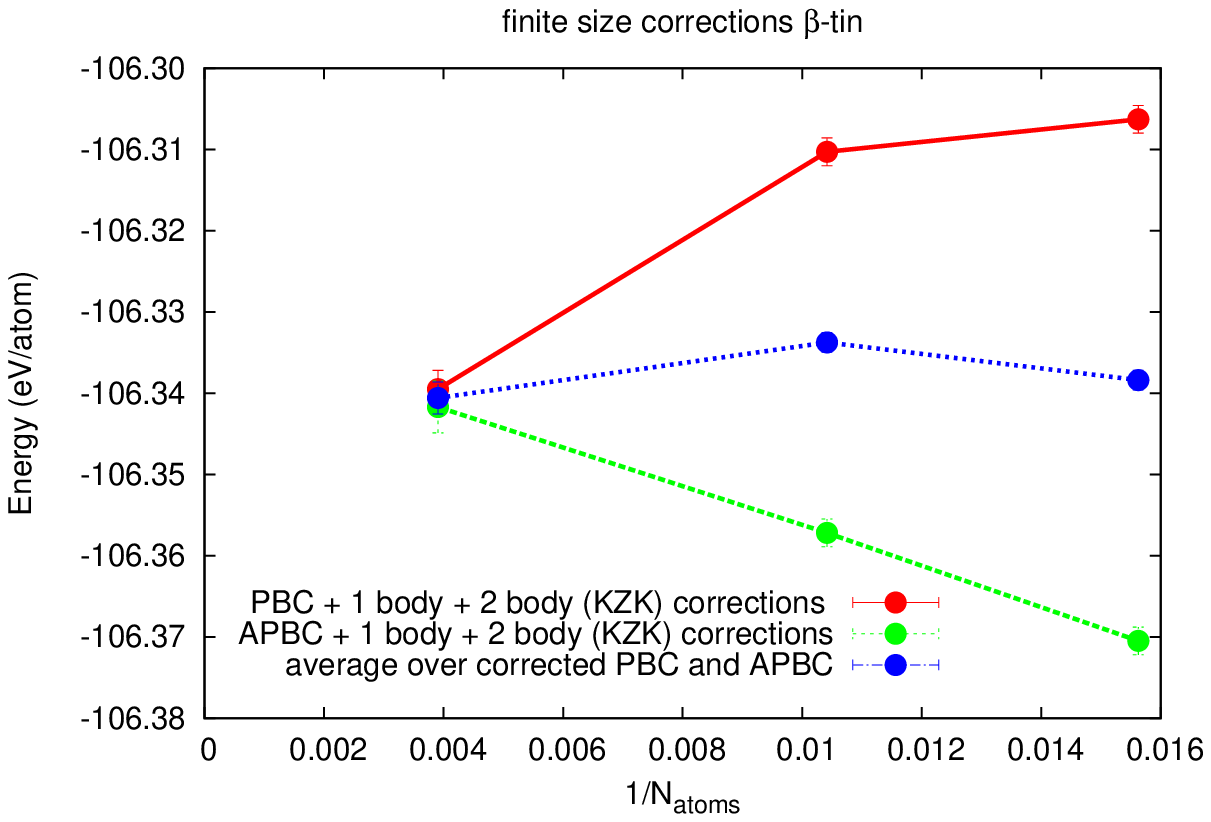}
\includegraphics[width=1.0\columnwidth]{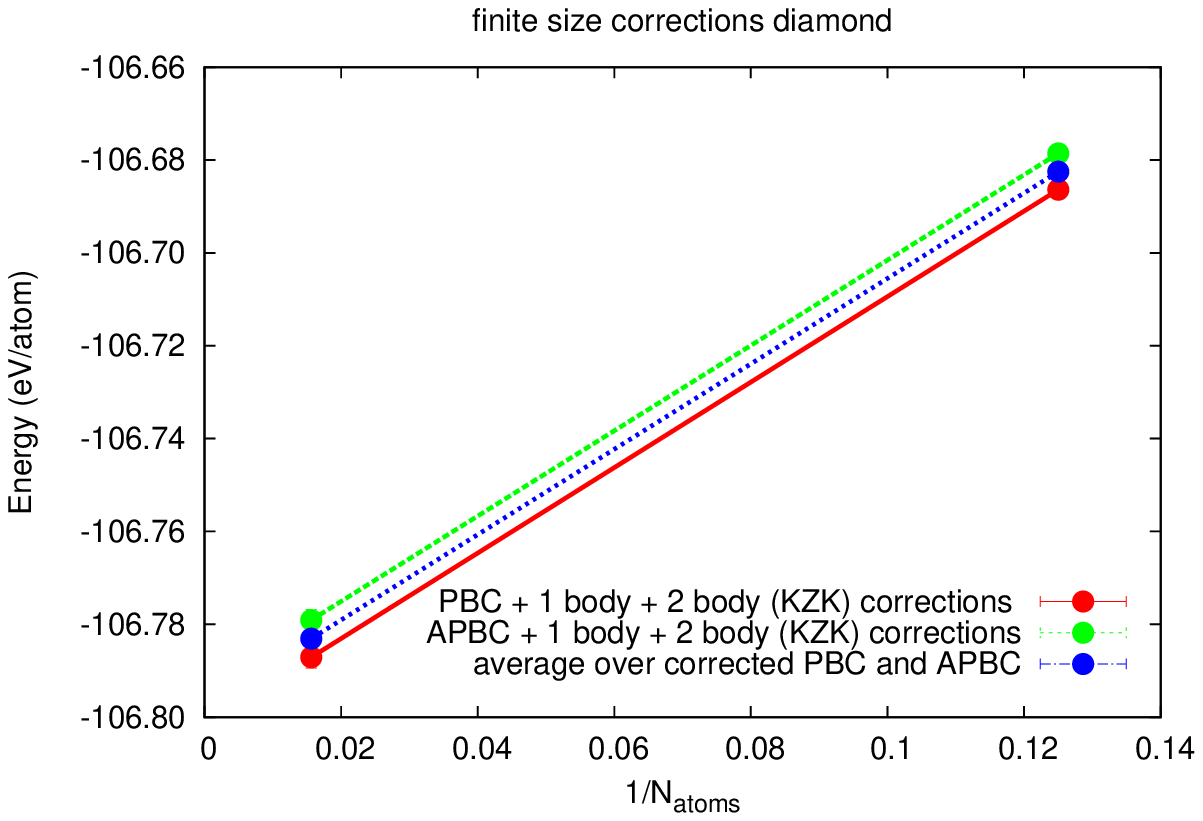}
\caption{
\label{size_plot}
Finite size extrapolation for the $\beta$-tin
  ($V =15$ \AA$^3$/Si, c/a=0.55)  and the diamond
  ($V =19.949$ \AA$^3$/Si)  phases in the
  upper and lower panel, respectively. The data points correspond to
  the values reported in Tab.~\ref{size}, obtained with the
  Trail-Needs pseudopotentials. In the plot the energy per
  atom is shown as a function of $1/N_\textrm{atoms}$, where $N_\textrm{atoms}$ is the number of Si
  in the crystal supercell. Note the slopes for a given boundary
  condition (either PBC or APBC) in the two cases differ by
  a factor of 10, making the extrapolation much harder for the
  metallic phase.
}
\end{figure}

For alleviating the effects of the one-body terms, we performed VMC
and LRDMC calculations averaging the KZK corrected energies over the 
two most symmetric points ($\Gamma$ and $M$).
Previous DMC calculations\cite{alfe, cyrus} were performed only at the 
$\Gamma $ point, leading to more pronounced size effects. 
In Tab.~\ref{size} we report the energy for different sizes together
with KZK corrections, while in Fig.~\ref{size_plot} we show how
important is to average over the $\Gamma$ (PBC) and $M$ (APBC) points
to reduce considerably the error in the finite size
extrapolation\footnote{It is clear that with the PBC+APBC average of
  the KZK corrected energies we
  can reduce the error on the final transition pressure well below the
  $20$ meV/atom threshold reached in Ref.~\onlinecite{cyrus}.}.
By averaging over PBC and APBC boundary conditions
we reach an accuracy of 5 meV/atom with 64 atoms, 
well below the magnitude of other
systematic errors, as we will see in Sec.~\ref{corrections}. Therefore, at
variance with Ref.~\onlinecite{cyrus}, where a single k-point was
adopted in the size extrapolations, the finite size bias is not
the largest error in our calculations.

\begin{table*} 
\begin{tabular}{|l|c|c|c|c|c|c|}
\hline
$N_\textrm{atoms}$    & $E^\textrm{QMC}_\textrm{PBC} +
 \Delta^\textrm{KZK}_\textrm{PBC}$ &
 $\Delta^\textrm{KZK}_\textrm{PBC}$  &
 $E^\textrm{QMC}_\textrm{APBC}+\Delta^\textrm{KZK}_\textrm{APBC}$  &
 $\Delta^\textrm{KZK}_\textrm{APBC}$  &
 $(E^\textrm{QMC}_\textrm{PBC} + E^\textrm{QMC}_\textrm{APBC} +
 \Delta^\textrm{KZK}_\textrm{PBC} + \Delta^\textrm{KZK}_\textrm{APBC})/2 $ &
 $(\Delta^\textrm{KZK}_\textrm{PBC} +\Delta^\textrm{KZK}_\textrm{APBC})/2$  \\  
\hline
64  $\beta-$tin &   -106.3063(17)  &   -.0654   & -106.3705(17)  & .2214   & -106.3384(12)  &   .07799  \\
\hline
96 $\beta-$tin &  -106.3103(17)  &   -0.0584  &  -106.3572(20)  &  0.0865   & -106.3338(13)   &  0.01409 \\
\hline
256 $\beta-$tin$^a$ & -106.3395(23) &   -.0273  &  -106.3417(32) &  .08855   & -106.3406(22) &  .03064 \\
\hline
64 diamond  &   -106.7871(23) &   -0.0329  &    -106.7791(23)  &   -0.0044   &  -106.7831(16) & -0.0187 \\
\hline
8 diamond  &   -106.6864(14) &  -.6700 &     -106.6786(14)  &   -.6632   & -106.6825(10)   & -0.6666 \\
\hline
\end{tabular}
\caption{  
\label{size}
Energy per atom (in eV) for various system sizes
in the metallic $\beta-$tin ($V=15$ \AA$^3$/Si, c/a=0.55) 
and insulating diamond ($V =19.949$ \AA$^3$/Si) phases with the
Trail-Needs pseudopotentials.
$E^\textrm{QMC}_\textrm{(A)PBC}$ is the variational QMC energy with
the given PBC or APBC boundary
conditions. $\Delta^\textrm{KZK}_\textrm{(A)PBC}$
is the total one-body and two-body correction computed by means of
the KZK energy functional, with the same boundary conditions.}
\begin{description}
\item{$a$} Corrected by $-0.0082(13)$eV/atom to take into account that in this 
case for the long distance tail of the Jastrow  
a few variational parameters ansatz was adopted, as described 
in Sec.~\ref{wave_function}. 
The same form was used in the $64-$Si case to estimate this
correction.
\end{description}
\end{table*}

\section{Impact of various approximations}
\label{corrections}
\subsection{Pseudopotential approximation}
\label{pseudo_approx}
 
The change in Silicon coordination number (from $4$ to $6$), related to 
the structural transition 
from diamond to $\beta$-tin geometries, may affect the transferability of Si PP's in the two phases.

\begin{figure}
\centering
\includegraphics[width=1.0\columnwidth]{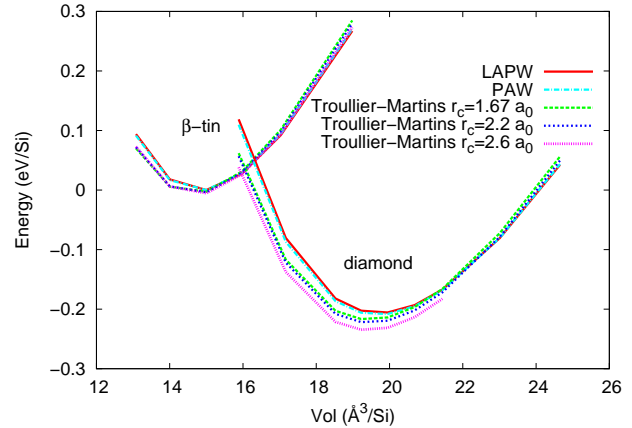}
\caption{
\label{comp}
EOS obtained by DFT LDA calculations with PWSCF and LAPW methods.
The PP's in the PW formalism have been generated by the
Troullier-Martins scheme with the cutoff radius $r_c$ reported in the
figure key. The zero of the energy has been chosen to be the minimum
of the $\beta$-tin EOS, corresponding to the volume $V = 15$
\AA$^3$/Si. This choice helps the comparison among the EOS curves.
Note that the PWSCF calculations done with the PAW pseudopotential are on top
of the all-electron LAPW points.
}
\end{figure}

We verified the impact of PP's on our final results, by estimating the
transition pressure at the DFT-LDA level with different norm conserving
PP's and with the projector augmented-wave (PAW) method\cite{blochl,kresse}.
\footnote{ The parameters of the PAW data set are  
$r_s=2.1$, $r_p=2.1$, $r_d=2.0$  ($d$ is local).
The nonlocal channels are fitted at the eigenvalue and at
$6.0$ Ry for $s$ and $p$. The unbound $d$ channel is at $\epsilon=-0.3$ Ry.
}
The results
have been compared with all electron calculations obtained with the
Wien2k\cite{wien2k} code. In Fig.~\ref{comp}, we report the EOS obtained
with norm conserving PP's generated by the Troullier-Martins method 
with scalar relativistic corrections and different cutoff radii $r_c$,
the data obtained by the PAW method, and the reference all-electron 
results.
The PP DFT calculations have been done with the PWSCF code\cite{pwscf}.
We worked with a plane wave cutoff of 50 Ry and a charge density cutoff
of 200 Ry. The number of non-equivalent k-points in the Brillouin zone
is 160 for the $\beta$-tin phase, and 80 for the diamond
structure. We checked that those parameters give converged DFT
results with a Gaussian broadening of 0.01 Ry of the Fermi surface.\cite{mp}
On the other hand, the Wien2k calculations are PP error
free, and therefore they can be used to check the PP accuracy.
They have been performed with an equivalent Brillouin zone integration over a $17
\times 17 \times 17$ k-point mesh, a muffin-tin radius
$R_\textrm{MT}=1.90 ~ a_0$ ($a_0$ is the Bohr radius), and a plane wave cutoff $k_\textrm{max}$
given by $R_\textrm{MT} k_\textrm{max} = 10$. These parameters give
converged results. If the error from the PP
approximation were negligible, all EOS curves would superimpose on
each other. 
Instead, Fig.~\ref{comp} shows that the EOS differ significantly. As
reported in Tab.~\ref{pseudo_press_dft}, the
transition pressure seems to converge with respect to the
Troullier-Martins core radius as it gets small.  Its main effect is to shifts the relative position between the
diamond and $\beta$-tin EOS branches. The EOS from PP's calculations are however  different from the all-electron one, even for the smallest $r_c$. Our results clearly show how the prediction of properties under pressure is affected by the PP approximations. The best choice in the DFT framework is to use the PAW pseudopotentials, which give both the transition pressure and EOS very close to the all-electron results.
\begin{table} 
\begin{tabular}{|r|c|}
\hline
DFT method       &  $p_t$(GPa)    \\
\hline     
LDA LAPW                                      &  7.12     \\
LDA PW PAW                                       &  7.21  \\
LDA PW Troullier-Martins $r_c=1.67 ~a_0$       & 7.44  \\
LDA PW Troullier-Martins $r_c=2.2 ~a_0$       &  7.65  \\   
LDA PW Troullier-Martins $r_c=2.6 ~a_0$       &  8.27  \\  
\hline
\end{tabular}
\caption{
\label{pseudo_press_dft}
 DFT transition pressures for different pseudopotentials in PWSCF
 calculations and all-electron Wien2K LAPW calculations
 done with the LDA functional. Note the convergence of the
 Troullier-Martins pseudopotential with respect to the core radius $r_c$.
By using soft pseudopotentials, the error could be of $1$ GPa on the
final transition pressure. The best DFT pseudopotential is the PAW one,
with a final transition pressure within 0.1 GPa from the all-electron result.
}
\end{table}

At the QMC level additional errors may came  from  the lack of a consistent method  to generate 
PP's from the corresponding correlated QMC calculation for an isolated
atom \footnote{ Early attempts to generate PP's at the QMC level was
  done in Ref.~\onlinecite{alcioli}}. A direct evaluation of the
core-valence interaction was attempted in
Ref.\onlinecite{alfe}. 

In Tab.~\ref{pseudo_press} we report our VMC and LRDMC results 
using  Hartree-Fock, energy adjusted and LDA generated pseudopotentials. 
The transition pressures reported in Tab.~\ref{pseudo_press} are evaluated by
computing the energy of the diamond and $\beta$-tin phase at the
volume per atom of $13.08$\AA$^3$  and $20.69$ \AA$^3$ respectively, and
assuming that the curvature of the corresponding EOS is the same as
the one computed for the Trail-Needs pseudopotentials.
\begin{table} 
\begin{tabular}{|r|c|c|}
\hline
pseudopotential       &  VMC   & LRDMC  \\
\hline     
HF Trail-Needs                                           & 15.48(6)    & 16.65(15) \\
energy adjusted Burkatzki-Filippi-Dolg     & 15.80(6) &  16.50(12)  \\
LDA Troullier-Martins $r_c=1.67 ~a_0$      & 15.39(18)   &  16.83(19) \\
LDA Troullier-Martins $r_c=2.1 ~a_0$        & 15.08(18)  &  16.11(19)  \\  
\hline
\end{tabular}
\caption{
\label{pseudo_press}
 Transition pressures for different pseudopotentials in VMC
  and LRDMC calculations. Calculations are done at volume per atom of
  $13.08$\AA$^3$  and $19.95$\AA$^3$ for the diamond and $\beta$-tin phase
  respectively, and the Maxwell construction done assuming for the EOS
  the same curvature as in the Trail-Needs case, fully resolved with
  respect to its volume dependence. $r_c$ is the cutoff radius in the
  pseudopotential generation scheme. The factor used to convert energy differences into pressures is 
  26.812 GPa/eV.
}
\end{table}

Results in  Fig.~\ref{comp}  and Tab.~\ref{pseudo_press}  
clearly shows  the interaction between core and valence 
electrons  cannot be approximated  by a rigid shift in energy
( as usually assumed  estimating the effect of PP on the transition pressure).

Core-valence interactions accounts for a correction of $1.2 \pm 0.6$ GPa and  
were calculated in Ref.~\onlinecite{alfe}. 
In order to improve the accuracy of this correction 
an all electron calculation is required.
At present this is almost impossible within QMC, and
therefore an uncertainty of at least  1 GPa coming directly from the PP
approximation is unavoidable in our QMC findings.

\subsection{Phonons and temperature effects}
\label{finitet_approx}

The inclusion of finite temperature and zero point motion effects is crucial
for a direct comparison of our results with finite temperature experiments 
(usually performed at  room temperature). Both experiments \cite{exp7}
and theory \cite{temp_dep} indicate that temperature corrections
induce a positive shift  to the critical equilibrium line. On the
other hand, a further shift in pressure is induced by the inclusion of
zero point motion effects. Phonon dispersions are in fact different in
the two phases and zero point motion effects do not compensate.   
\begin{table*} 
\begin{tabular}{lcccccc}
\hline
\textrm{V}  (\AA$^3$/Si)     & 0K  &  100K & 300K & 500K &  700K & 1000K \\
\hline
   & & & Diamond phase & & &  \\ 
  15.882  &  0.077/0.073 &   0.073/0.069 &  0.042/0.039 & -0.012/-0.016 &  -0.084/-0.090 & -0.218/-0.226 \\
  17.169 &  0.069/0.068  & 0.067/0.066 &  0.040/0.041 & -0.012/-0.011 & -0.083/-0.081 &  -0.215/-0.212 \\
  18.524 &  0.066/0.064 &  0.064/0.063 &   0.039/0.038 &  -0.012/-0.014 &   -0.082/-0.084 & -0.213/-0.216 \\
  19.228 &  0.064/0.062 &   0.062/0.061 &  0.038/0.036 &  -0.014/-0.016 &   -0.084/-0.087 &   -0.216  -0.220 \\
  19.949  & 0.062/0.060  &  0.060/0.059 &   0.036/0.034 & -0.017/-0.019 &  -0.088/-0.091 & -0.221/-0.225 \\
  20.687 &  0.060/0.058 &  0.058/0.057 &   0.033/0.031 &  -0.020/-0.022  & -0.092/-0.095 & -0.226/-0.231 \\
  21.444 &  0.058/0.056 &  0.056/0.055 &  0.031/0.029 & -0.023/-0.026 &  -0.096/-0.101  & -0.233/-0.238 \\
  23.013 &  0.054/0.052 &  0.052/0.051 &  0.025/0.023 & -0.032/0.034 & -0.108/0.112 & -0.249/0.254 \\
  24.656 &  0.050/0.048 &  0.048/0.047 &  0.019/0.017 & -0.040/0.044 & -0.120/0.124 & -0.265/0.271  \\
   & & & $\beta-$Sn phase & & & \\
  13.081 &  0.055/0.053 &  0.053/0.051 &  0.025/0.022 & -0.033/0.036 & -0.110/0.114 & -0.251/0.257  \\
  14.004 &  0.049/0.047 &  0.048/0.046 &  0.017/0.014 & -0.044/0.048 & -0.125/0.130 & -0.272/0.280  \\
  15.000 &  0.044/0.042 &  0.042/0.040 &  0.008/0.005 & -0.058/0.063 & -0.144/0.150 & -0.299/0.308  \\
  15.978 &  0.039/0.037 &  0.036/0.035 & -0.002/0.005 & -0.073/0.078 & -0.164/0.172 & -0.328/0.339  \\
  17.031  &  0.034/0.032 &   0.030/0.028 &  -0.014/0.019 &  -0.093/0.100 &  -0.192/0.202 &  -0.368/0.382 \\
  18.984 &   0.026/0.025 &   0.020/0.019 &  -0.037/0.040 &  -0.129/0.135 &  -0.242/0.250 &  -0.438/0.450 \\
\hline
\end{tabular}
\caption{  Finite temperature corrections to the free energy (eV/atom)
within PBE/LDA DFT theory, as explained in the text.  
\label{finitettab}}
\end{table*}

Finite temperature effects and zero point motion energies are included
in our estimate of the transition pressure, by calculating phonon
dispersions for the $\beta$-tin and diamond phase with the QE package.   
In order to study the convergence of our phonon calculations with the
k-point mesh we performed PWSCF runs 
with large plane wave cutoff (up to 100 Ry), accurate k-point sampling
(up to $1620$ inequivalent 
k-points for the $\beta-$tin phase and $480$ inequivalent k-points for
the diamond phase), as well as a small value of the Gaussian
broadening (width of 0.0001 Ry).  
We have used the PAW data set
as in the previous section for the LDA calculation, and an ultra-soft (US) PP 
generated with similar parameters and core radii for the
PBE calculation.

Following Ref.~\onlinecite{temp_dep}, we compute the harmonic
correction $\Delta F$  to the free energy per atom, 
 by using the phonon density of states $D(\omega)$
 available in QE after  Fourier interpolation of the phonon bands, 
(namely by using matdyn.x): 
\begin{equation}
\Delta F = kT \int\limits_0^\infty d\omega D(\omega)  \ln \left(2 \sinh \left( { \omega\over 2 kT}\right) \right)  \end{equation}
where $\hbar=1$ is assumed.
In using the above expression, one has to take into account that the 
total phonon density of states per atom is obviously normalized to 
\begin{equation} \label{sumdos}
 \int\limits_0^\infty d\omega D(\omega)=3
\end{equation}
as there are three phonon modes per atom in the thermodynamic limit.
Integrations were changed to summations over 
a  uniform mesh with high resolution ($1cm^{-1}$), and the original density 
of states was appropriately scaled to fulfill Eq.~\ref{sumdos}.
 
Free energy corrections   (see Tab.~\ref{finitettab}) 
are then added to 
a total energy zero temperature calculations.
In this way 
our calculation of the transition 
pressure, estimated by the Maxwell construction of the free energy 
curves,
is essentially free of systematic errors within the 
chosen DFT functional, as long as phonon anharmonic effects can be neglected 
in the low temperature regime.
This is a reasonable assumption below the melting temperature occurring at 
about $\simeq 1000K$. 

\begin{figure}
\centering
\includegraphics[width=1.0\columnwidth]{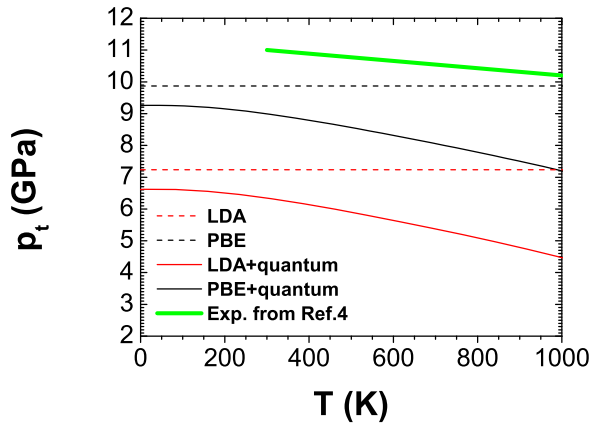}
\caption{
\label{ptdft}
Effect of harmonic quantum corrections on the transition pressure in Silicon
as a function of temperature. 
}
\end{figure}

To test the impact of the exchange and correlation functional on  quantum correction estimates, we have performed  calculations with both the PBE and LDA functionals. 
As shown in Fig.~\ref{ptdft},  although different functionals provide different transition pressures,
the corrections to the bare values are very similar and consistent within $0.2$ GPa and in fair agreement with experimental results.
The results  demonstrate that phonons are rather well described within DFT and these 
corrections are very reliable at least before the melting point.
Our results do not agree with  a  previous work\cite{temp_dep} on 
this subject, where the zero temperature quantum corrections were
underestimated by about a factor two, and  finite temperature corrections were larger by about a factor three. 
PP used in Ref.\onlinecite{temp_dep} is no more available and we were not able to reproduce the quoted results. Presently the reason for this discrepancy
is not clear.
  
With  new available PP's, our temperature corrections 
 are very well converged, and  appear in  reasonable agreement with recent experimental data (see Fig.~\ref{ptdft}).

\begin{figure}
\centering
\includegraphics[width=1.0\columnwidth]{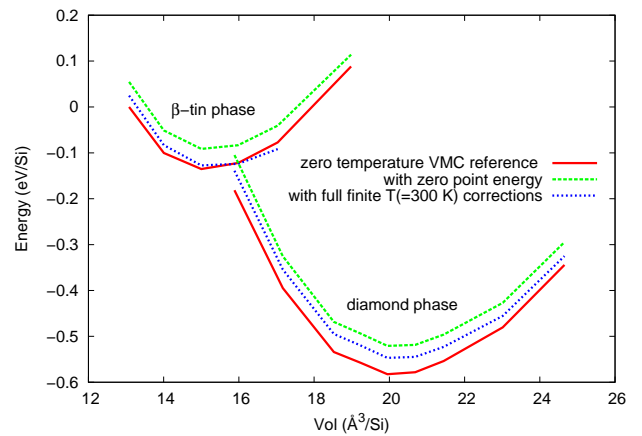}
\caption{
\label{finite_temp_corr}
The corrected equation of state once the zero point energy and the
temperature effects are added (estimated at the DFT-PBE level). The
reference is the VMC calculation reported in Fig.~\ref{vmc}, and here
the zero of energy is taken as the VMC value at $V=13.08$ \AA$^3$/Si.  
}
\end{figure}

By 
using  DFT-PBE free energy corrections to  
 VMC total energies, we
obtain the corrected VMC curve reported in Fig.~\ref{finite_temp_corr},
and the corrected transition pressures in Tab.~\ref{corr_press}.
\begin{table} 
\begin{tabular}{|r|c|c|}
\hline
                                         &  VMC $p_t$ & LRDMC $p_t$   \\
\hline     
Born-Oppenheimer                                   & 15.48(6)   & 16.65(15) \\
zero point motion                                     & 14.84(6)  &  15.71(14) \\
$T=300$ K                                         & 14.53(6)    & 15.74(17) \\
\hline
\end{tabular}
\caption{
\label{corr_press}
 Transition pressures $p_t$ in VMC and LRDMC calculations with quantum and
  temperature effects. The values are obtained by a Maxwell
  construction on a cubic fitting of the EOS, corrected at each volume
  by the DFT-PBE estimates. 
}
\end{table}
As one can note, the zero point energy for the diamond phase is larger than
the one for the $\beta$-tin phase by $\approx 0.2$ eV, which decreases
the transition pressure by $0.65$ GPa. The finite temperature
correction is negative, and its absolute value is larger for the $\beta$-tin
phase. Therefore, a further reduction of $0.30$ GPa is obtained at
$300$K, which implies that the total correction at room
temperature is $0.95$ GPa, a smaller value than the one
estimated in Ref.~\onlinecite{temp_dep}. At the LRDMC level, we obtain
roughly the same total correction ($0.91$ GPa), although it is more
difficult to discriminate the temperature effect, as the statistical error is
larger (see Tab.~\ref{corr_press}).

\subsection{Kleinman-Bylander approximation}
\label{kb_approx}
The matrix elements of the non-local PP can be evaluated 
either by direct numerical Gauss-Hermite (GH) integration over the polar coordinates
or by using the Kleinman-Bylander (KB) approximation\cite{kb_approx,kb_approx2}.
The KB approach is a rather general concept, and it is applied in DFT calculations
to make the calculation of the PP operator more efficient.
In particular, in the plane-wave formalism the generated PP
is conveniently expressed in the plane-wave basis set.
On the other hand, in the QMC framework, one usually works 
in the
coordinate representation where electron positions and spins are given, and this
makes the KB construction hard to
 implement numerically.
Consequently, the pseudopotentials used in QMC are usually generated
in the so called semilocal form (local + non local part), and computed by
performing  a random integration over their angular components\cite{fahy,mitas,lrdmc}.
However, in previous QMC calculations of the Si diamond-to-$\beta$-tin transition\cite{alfe,cyrus}  
the determinantal part has been generated from plane-wave DFT calculations, where the KB approximation
was used\cite{cyrusp}  to represent a pseudopotential originally written in a semilocal form.
This procedure could lead in principle to a poorer form of the variational wave function
in the proximity of the core, where the non-local PP is mostly localized, with an
impact on its nodal structure, and therefore a larger FN error.

\begin{table*}[!ht]
\begin{tabular}{|l|c|c|c|c|}
\hline
Quantity    & KB  & GH  & present  $n=24$  &  present $n=10$ \\  
\hline
Total energy  & -102.85692  & -102.84036 &  -102.83975  & -102.83737 \\
\hline
Kinetic  energy & 50.95461   &  50.83712  &   50.83563 & 50.83195 \\
\hline
Non-local pseudo &  22.13269   &  22.45092  &  22.45256 &  22.45398 \\
\hline
\end{tabular}
\caption{
\label{si1tab}
Comparison of LDA-DFT energies (eV/Si) obtained within the KB approximation and 
with the GH numerical integration over the polar coordinates of non-local PP matrix elements. 
We report the total energy, the kinetic energy and the non-local contribution of the PP 
for a system of 8 Si with anti periodic boundary conditions ($M-$ point) in the $\beta$-tin phase with $c/a=0.54$ and volume $13.081$\AA$^3/Si$ and Trail-Needs pseudopotentials.
Calculations within the KB approximation are performed with {\it PWscf} code. 
Plane-wave calculations with a GH numerical integration are done using {\it Qbox} code.
We report also the energies obtained for the same system using the Gaussian basis set implementation of the LDA-DFT method
coded in the {\it TurboRVB} package (referred to as ``present'' in the table). $n$ indicates the 
number of Gaussians for single-particle orbitals with the $s$ symmetry. 
}
\end{table*}

\begin{table*}[!ht]
\begin{tabular}{|l|c|c|c|c|}
\hline
Quantity    & KB  & GH & present  $n=24$  &  present $n=10$ \\  
\hline
Total energy  & -105.32986  & -105.31911   &  -105.31880  &   -105.31302 \\
\hline
Kinetic  energy  &  44.0823   & 43.96216 &   43.96152   &  43.95454    \\
\hline
Non local pseudo & 21.43672   &  21.79496 &   21.79506 &   21.79468  \\
\hline
\end{tabular}
\caption{
\label{si2tab}
Same as in Tab.~\ref{si1tab} but for a system of 8 Si with anti periodic boundary conditions 
in the diamond phase at volume $20.036$\AA$^3/Si$.
} 
\end{table*}

To investigate the impact of KB approach, we compared the total energy, the kinetic energy, 
and the non-local term of the PP, obtained from a single M-point DFT calculation with and without the KB approximation.  
The results are shown in Tab.~\ref{si1tab} for a system of $8$ Si in the 
$\beta$-tin phase with $c/a=0.54$ and volume $13.081$\AA$^3/Si$, and in the 
diamond phase with volume $20.036$\AA$^3/Si$. 
Calculations within the KB approximation are done 
using the {\it PWscf} DFT implementation\cite{pwscf}. We use the {\it Qbox} code\cite{qbox} for performing 
plane-wave calculations with Gauss-Hermite integration. An energy cut-off of $100$ Ry was used 
in all the plane-wave calculations. The exchange and correlation energy was described in the LDA
by the Perdew-Zunger functional\cite{perdew}.
The energies reported in Tab.~\ref{si1tab} and Tab.~\ref{si2tab} clearly show 
that the use of KB approximation causes an error of 0.3 eV/Si when evaluating 
the contribution from the non-local term of the PP.
 This error cancels out in the energy difference between the two phases,
leaving the DFT results unbiased by this kind of approximation. 
In principle, our analysis cannot exclude that 
the nodal structure of the DFT generated wave function is unaffected close to the core,
since there is a significant KB error 
in the PP contribution of the total energy.
In practice, the close agreement between our LRDMC results, unaffected
by the KB approximation, and the ones in
Ref.~\onlinecite{cyrus}, where this approximation has been used, shows
that the cancellation of KB errors applies also in QMC
calculations and leads to unbiased energy differences.

Our results are unaffected by the KB approximation by construction,
as the DFT code implemented in the {\it TurboRVB} package uses the
exact GH integration of the semilocal pseudopotentials. Moreover, 
the DFT results shown in Tab.~\ref{si1tab} and Tab.~\ref{si2tab} are a further validation 
of the convergence of our periodic Gaussian basis set. Indeed, we found a perfect 
agreement between the energies obtained with plane-wave calculations (performed by means of the {\it Qbox} code \cite{qbox}
and without the KB approximation) and the ones obtained with our Gaussian basis set implementation 
in the {\it TurboRVB} package\cite{turboRVB}.

\section{Results}
\label{results}

In this section we present our QMC results for the transition
pressure, performed with the Trail-Needs pseudopotentials.
We first validate the quality of our variational wave function
in both the $\beta$-tin and diamond phases.
In the $\beta$-tin phase, 
we compute the $c/a$ ratio in the proximity of
the critical pressure (at volume $V=15$ \AA$^3$/atom), where this lattice
parameter is experimentally known. We evaluate also the 
bulk modulus and other structural properties
from the fit of the equation of state $E=E(V)$, and compare them with
former experimental and theoretical values. Finally, we compute the transition
pressure by performing the Maxwell construction on the EOS of both phases.

\begin{figure}
\centering
\includegraphics[width=1.0\columnwidth]{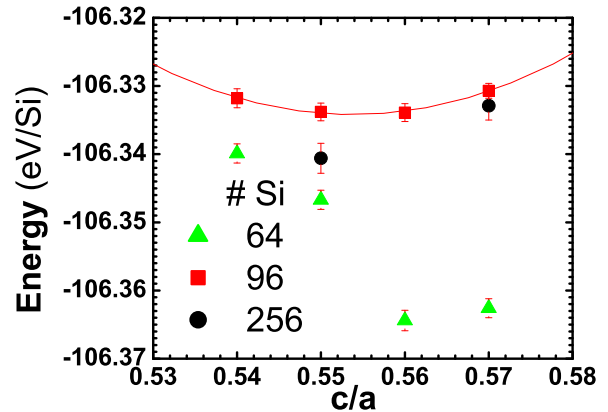}
\caption{
\label{evscsua}
Energy vs $c/a$ in the $\beta-$tin phase at  
fixed volume $15$\AA$^3/Si$ for various sizes. 
FS corrections from the KZK extrapolation are included and energies are averaged over the  
two most symmetric points ($\Gamma$ and $M$) of the Brillouin zone. In the largest size the Jastrow factor was parameterized 
at large distance according to Eq.(\ref{parametrization}) 
and the error of the restricted variational 
freedom ($\simeq -0.0082$ eV/atom) was estimated  with the same calculation 
for 64 Si atoms. 
}
\end{figure}

In Fig.~\ref{evscsua} we report the VMC energies as a function of the
$c/a$ ratio for the $\beta$-tin phase at different supercell size
($64$, $96$, $256$ Si atoms) and fixed volume $V=15$ \AA$^3$/atom. All the
energies include FS corrections within the KZK scheme, as reported in
Subsec.~\ref{finite_size}. The results in Fig.~\ref{evscsua} are a first test of the
quality of our WF for describing the $\beta$-tin phase. With the
Jastrow optimization  we reproduce rather well 
the experimental value $c/a=0.554$. Therefore for 
all the following VMC and LRDMC calculations we fix 
the $c/a$ ratio to the value $0.55$. 

\begin{figure}
\centering
\includegraphics[width=1.0\columnwidth]{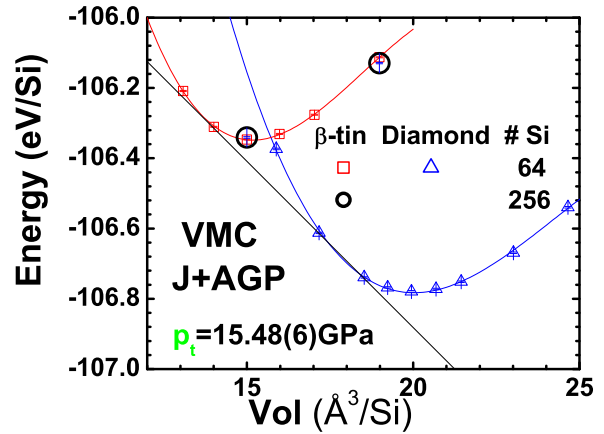}
\caption{
\label{vmc}
\small{ 
Equation of state and the metal-insulator transition pressure in bulk Silicon 
obtained by the VMC technique after optimizing a Jastrow factor in a 
localized basis set containing $2s$ and $2p$ Gaussian orbitals per Si. 
The energies are corrected using the KZK correction scheme.\cite{KZK}
In this way finite size effects of the transition pressure appears to 
be small. 
The lattice value $c/a=0.55$ is used for the $\beta-$tin phase.
The data plotted here are not yet corrected for the quantum and temperature effects.
See Tab.~\ref{sifinal} for the final pressure which includes zero point motion and thermal contributions.
 } }
\end{figure}

\begin{figure}
\centering
\includegraphics[width=1.0\columnwidth]{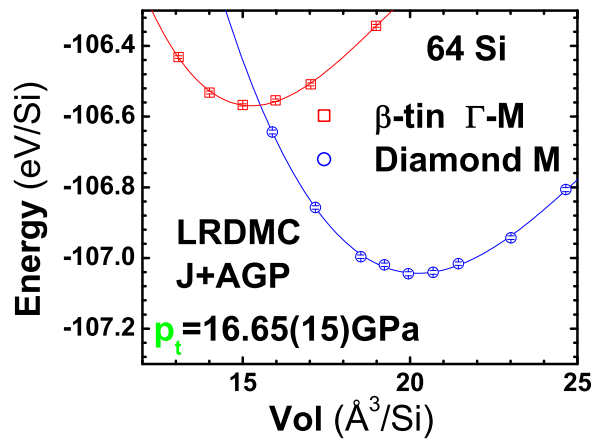}
\caption{
\label{lrdmc}
\small{ 
The same as in Fig.~\ref{vmc} but for the LRDMC calculations.
} }
\end{figure}

In Fig.~\ref{vmc} we reported the VMC results for the energy $E(V)$ as a function of the volume $V$.
All the energies in the Figure include FS corrections using the KZK
scheme. Although FS effects are more pronounced in the metallic phase,
the results for the $\beta$-tin phase clearly show that the FS errors are
under control. In fact we find that the energies for the $256$ atoms
supercell fall on the top of the data for the $64$ atoms calculations.  
LRDMC energies are shown in Fig.~\ref{lrdmc} for $64$ atoms supercell.
VMC and LRDMC EOS are fitted using a cubic polynomial
function.
The results for bulk properties of the diamond phase, reported in
Tab.~\ref{sistruct}, are in very good
agreement with the experimental values, whereas the ones for the 
$\beta-$tin structure compares well with previous QMC data. 
Note also that there exists a sizable zero point motion correction to both 
the equilibrium volume and the bulk modulus, that has not been taken 
into account so far in previous works.
We have estimated these corrections by adding to the EOS the 
zero temperature quantum corrections evaluated within the harmonic
approximation and the PBE functional, as explained in
Subsec.~\ref{finitet_approx}. 

\begin{table} 

\begin{tabular}{|l|c|c|c|}
\hline
diamond phase & $\textrm{V}_\textrm{eq}$  (\AA$^3$/Si) &
$\textrm{E}_\textrm{cohesive}$ (eV/Si) & B (GPa) \\
\hline     
LDA                                        &    19.77
&   5.29  & 95.75 \\
PBE                                         &   20.42
& 4.62 & 89.4 \\
VMC                                       & 20.124$\pm$0.036\footnotemark[1] 
&  4.6003$\pm$0.0015\footnotemark[2] & 102.3$\pm$1.4\footnotemark[3] \\
LRDMC                                   & 20.33$\pm$0.1\footnotemark[1]
& 4.6650$\pm$0.003\footnotemark[2]  & 95.78$\pm$3\footnotemark[3] \\
DMC (Ref.~\onlinecite{alfe})   & 20.21$\pm$ 0.03\footnotemark[1]  
& 4.62 $\pm$ 0.01 \footnotemark[2] & 101.4$\pm$10 \footnotemark[3] \\
DMC (Ref.~\onlinecite{cyrus}) &  20.08$\pm$0.05 \footnotemark[1] 
& - & 98$\pm$7 \footnotemark[3]   \\
Exp. &  20.0\footnotemark[4]         
& 4.62$\pm$0.08\footnotemark[5]  & 99\footnotemark[4] \\
\hline
$\beta-$Sn phase  & $\textrm{V}_\textrm{eq}$  (\AA$^3$/Si) &
$\textrm{E}_\textrm{cohesive}$ (eV/Si) & B (GPa) \\
\hline
LDA &  14.92 & 5.10 &  115.4 \\
PBE &  15.45 & 4.29 &  110.7 \\
VMC &  15.25$\pm$ 0.05 \footnotemark[6] & 4.186 $\pm$ 0.0013 \footnotemark[7] & 119.7 $\pm$ 3.5\footnotemark[8] \\ 
LRDMC & 15.34 $\pm$ 0.16 \footnotemark[6] & 4.211 $\pm$ 0.0024 \footnotemark[7]
  & 111.3 $\pm$ 8.3 \footnotemark[8] \\  
DMC (Ref.~\onlinecite{cyrus}) &  15.31$\pm$0.2 \footnotemark[6] 
& - & 98.6$\pm$12 \footnotemark[8]   \\
\hline
\end{tabular}
\caption{ 
\label{sistruct}
Comparison of present numerical results with previous QMC data
and available experiments for the equilibrium properties of
Silicon in the diamond structure: equilibrium volume ($\textrm{V}_\textrm{eq}$), cohesive
energy ($\textrm{E}_\textrm{cohesive}$), and bulk modulus (B) are reported.
They were estimated by a cubic interpolation of the 
$E(V)$  points. 
The DFT pseudopotentials used here includes non linear core polarization
terms and scalar relativistic corrections and are in 
consistent agreement with corresponding all-electron calculations.
For the DFT methods the cohesive energy is given by the magnetic solution 
of the single atom, whereas quantum corrections are included according to 
Tab.(\ref{finitettab}).
All corrections to QMC data  are estimated  within the DFT-PBE 
functional.
We have also checked that standard LDA functional provides similar
corrections. 
} 
\footnotetext[1]{ Corrected by 0.1\AA $^3$/atom to take into account the zero point motion.} 
\footnotetext[2]{ Corrected by -0.06eV/atom to take into account the zero point motion.}
\footnotetext[3]{ Corrected by -1.6GPa to take into account the  zero point motion. 
VMC and LRDMC are further corrected by $-10$ GPa to take into account non cubic terms in the
interpolation within the equilibrium volume range  $ 13
\leftrightarrow 19$ ($16 \leftrightarrow  25$) \AA$^3/{\rm atom}$ in the $\beta-$tin (diamond) phase.} 
\footnotetext[4] {Taken from Ref.~\onlinecite{Siexp1}}
\footnotetext[5] {Taken from Ref.~\onlinecite{Siexp}}
\footnotetext[6] { Corrected by 0.11\AA $^3$/atom to take into account the zero point motion. }
\footnotetext[7]{ Corrected by -0.04eV/atom to take into account the zero point motion.}
\footnotetext[8]{ Corrected by -2.8GPa to take into account the  zero point motion. VMC and LRDMC are further corrected by -5.6Gpa analogously to the diamond phase.}
\end{table}

The critical pressure of the diamond to $\beta$-tin transition is reported in
Tab.~\ref{sifinal}. VMC calculations give a raw $p_t$ of
$15.48(6)$ GPa, LRDMC data give $16.65(15)$ GPa.

The inclusion of zero point motion, finite temperature and core-valence contributions bring the transition pressure to $13.33(70)$ GPa (VMC) and $14.50(70)$ GPa (LRDMC), for a total final shift  of $-2.15$ GPa (Tab.~\ref{sifinal}). Zero point motion and thermal corrections at $300$ K amount to $-0.65$ GPa and to $-0.3$ GPa respectively (they are estimated performing a PBE phonons calculations, as explained in Subsec.~\ref{finitet_approx}).
The core-valence interaction contribution is $-1.20 \pm 0.6$ GPa from Ref.~\onlinecite{alfe}, the same value has been used in Refs.~\onlinecite{cyrus,alfe,purwanto}. Contributions beyond frozen core approximation are already included at the DFT level trough non-linear core corrections in the PP.
We observe that previous calculations  \cite{alfe},\cite{purwanto}\cite{cyrus} consider a total correction to the raw data of $-2.5$ GPa, because of the different value of the zero point energy and finite temperature effects. All previous 
calculations should be increased by $0.35$ GPa, for accounting this difference.                           .
   
\begin{table} 
\begin{tabular}{|r|c|c|}
\hline
method       &  raw (GPa)  & corrected (GPa)  ($T=300$ K)  \\
\hline     
LDA                                        &  7.21  & 6.34 \\
PBE                                        &  9.87 &  8.99 \\
VMC                                       & 15.48$\pm$0.06   &  13.33$\pm$1.0  \\
LRDMC                                   & 16.65$\pm$0.15   &  14.50$\pm$1.0  \\  
DMC (Ref.~\onlinecite{alfe})   & 19.0 $\pm$ 0.5     &  16.5$\pm$0.5  \\
DMC (Ref.~\onlinecite{cyrus}) & 16.5$\pm$1.0       &  14.0 $\pm$ 1.0 \\
AFQMC (Ref.~\onlinecite{purwanto}) & 15.1 $\pm$ 0.3 & 12.6 $\pm$ 0.3   \\ 
Exp. & 10.0 - 12.5 & 10.0 - 12.5 \\ 
\hline
\end{tabular}
\caption{ 
\label{sifinal}
Zero temperature transition pressure in GPa obtained by a cubic interpolation of the
EOS. Comparison of the present numerical results with available experiments
and previous theoretical data. 
The corrected numerical QMC data are obtained after including
the zero point motion, finite temperature, and core-valence
contributions, which are not present in the raw data, as explained in
the text. DFT corrections include only the zero point motion and
finite temperature effects as they are performed with the non-linear core correction in the
pseudopotentials. In particular, the LDA results have been obtained by using a PAW PP,
while the PBE values come from an US PP.
 For the core-valence contributions 
 we have used the published DMC estimate
in Ref.\onlinecite{alfe}, that unfortunately is affected by a very large 
error bar, as discussed in the text.
The raw data and the corrections applied by other authors have been taken 
from the referenced papers. 
}
\end{table}

Both our corrected VMC and LRDMC values are above the experimental range,  
and remarkably the more accurate LRDMC method leads to a transition pressure 
$p_t$  larger than the VMC result. 

To further support the accuracy of our LRDMC calculations, we have 
systematically optimized the molecular orbitals by minimizing 
the VMC energy in presence of a Jastrow factor, starting from the LDA orbitals,
for a system of $8$ Silicon atoms (see Tab.\ref{optwf}).
The optimization of the molecular orbitals allows us to assert the impact of 
the wave function on the final LRDMC results.

\begin{table*}
\begin{tabular}{|l|c|c|}
\hline
 system  & DFT+J/LRDMC  &  FOPT/LRDMC  \\  
\hline
diamond \textrm{V}$=19.949$\AA$^3/Si$ PBC  & -106.0120(12)/-106.3064(27) &  -106.0493(10)/-106.3204(24)  \\
\hline
$\beta-$tin \textrm{V}$=13.081$\AA$^3/Si$ PBC  &  -106.5557(11)/-106.8765(28) &  -106.5890(10)/-106.8871(32)  \\
\hline
$\beta-$tin \textrm{V}$=13.081$\AA$^3/Si$ APBC  &  -103.7359(13)/-104.0720(41) &      -103.7617(9)/-104.0818(36)    \\
\hline
\end{tabular}
\caption{ 
\label{optwf}
Variational Monte Carlo/LRDMC  energy per atom (eV) 
for a system of 8 Si with periodic (PBC) 
and antiperiodic (APBC) boundary conditions obtained 
with the same basis: 8s6p4d  
for the Slater determinant, and 2s2p/1s1p for the Jastrow factor.
In the DFT+J case only the Jastrow factor (with no restriction to the exponents of the Gaussians) is optimized, while the 
determinantal part is the output of an LDA calculation in the 8s6p4d basis.
Conversely, in the fully optimized (FOPT) case    
the molecular orbitals are optimized together with the Jastrow factor, 
while keeping fixed the exponents of the 8s6p4d Gaussians to the even
tempered values discussed in the text (see Subsec.~\ref{wave_function}).
} 
\end{table*}

We have found that the LDA orbitals are a quite accurate starting point, 
but for the diamond at equilibrium geometry the total VMC energy decreases 
slightly more ($\simeq 0.01$eV/atom) than the one for the $\beta-$tin at high pressure, 
implying an increase in the transition pressure of about $0.4$ GPa at the VMC 
level. That is again consistent with the more accurate LRDMC
calculation, which is much less affected by the optimization 
of the wave function as clearly shown in Tab.~\ref{optwf}.
Altogether these results
point even more  clearly in the direction of a larger
zero temperature transition pressure.  

\section{Conclusions}
\label{conclusions}

We have performed DFT and QMC calculations for the diamond-to-$\beta-$tin
transition in Silicon. At the QMC level 
we have proven that it is possible to
accurately and efficiently 
describe correlation effects across the metal-insulator transition
by applying a relatively simple Jastrow factor to a DFT generated
Slater determinant.

We have shown that 
the estimation of the PP effect is one of the most delicate issues 
in any QMC calculation, and represents so far the 
most significant source of systematic error in the transition
pressure, amounting to about 1GPa, as described in Subsec.~\ref{pseudo_approx}.
The core-valence correlation correction strongly depends on the 
PP, and is almost unpredictable without doing the corresponding all-electron 
calculation. Moreover, the only known QMC estimate for this
correction ($-1.2$GPa), was done in 
Ref.~\onlinecite{alfe} for the Trail-Needs PP, and is affected by a large statistical error  ($\simeq 0.6$ GPa). 

In the present work we have also found a significant 
reduction of the phonon correction to the transition pressure-
from the quoted $-1.3$ GPa\cite{temp_dep} to our fully converged
LDA and PBE value of $-0.95$ GPa at $300$K.   
This shifts up all previous QMC $p_t$ estimates by $\approx 0.35$ GPa.
Thus, we arrive at the conclusion that \emph{all} Monte Carlo
findings published so far predict a transition pressure that is 
significantly larger than the one observed at room temperature, 
from $1/2$ GPa\cite{purwanto} to $4$ GPa\cite{alfe}  
above the experimental upper edge of $12.5$ GPa. 

Our VMC technique is in agreement with the value of the 
transition pressure recently reported by the AFQMC technique\cite{purwanto}.
A clear increase of the transition pressure by about 1GPa is obtained
when the 
accuracy of the calculation is improved by the LRDMC 
scheme.
Although LRDMC total energies are more accurate, it is in principle
possible that this technique  could lead to results worse than
the VMC ones 
in the estimation of the equation of state.
However we believe that this is quite unlikely, because 
DMC as well as LRDMC always improve the accuracy of the physical estimates, 
as long as they are applied to a reasonably good variational 
wave function. 

As a  further independent check 
that a more accurate transition pressure is larger than the VMC estimate,
we have  optimized the 
Slater determinant in presence  of the Jastrow factor for a small number 
of atoms, and found a consistent increase in the transition pressure.
Although we cannot estimate more accurately 
the nodal error -namely the exact ground state result for given pseudopotential- it looks plausible that by fully optimizing the wave function 
this error should decrease. Therefore, our orbital optimization 
should give at least the trend of the correction to the approximate VMC 
result. 

On the other hand, our LRDMC result is very close to the recent DMC
calculation performed by Hennig {\it et al.}\cite{cyrus}, where the same
pseudopotential was used, whereas the 
 DFT Slater determinant was obtained with the PBE functional and 
the KB approximation for the pseudopotential\cite{cyrusp}.
This agreement suggests also that the 
DMC/LRDMC 
technique is weakly dependent 
on the functional used to generate the DFT orbitals and should be considered rather accurate for 
a given pseudopotential.

To summarize, our best estimate of the transition pressure at room temperature is
14.5 GPa, with an uncertainty of 1 GPa coming mainly from the
pseudopotential approximation, as our finite-size extrapolation error
is definitely smaller. The discrepancy with respect to the
experimental values leads us to conclude that further work is necessary 
to determine the phase boundary of the metal-insulator transition in 
Silicon. On one hand, from the experimental point of view one should 
verify whether,  by removing the stress anisotropy in the experimental 
environment, the transition pressure 
can significantly increase and get closer to the QMC prediction, as suggested 
in Ref.~\onlinecite{cyrus}.  
On the other hand, in QMC calculations it should be worth defining  
consistent pseudopotentials, since we have seen that they can significantly 
affect the EOS at large pressure. So far there is in fact no accurate method 
to estimate the systematic error related to the pseudopotential approximation,
 since an all-electron calculation of bulk silicon is basically prohibitive within 
the QMC method. 
A first attempt along these lines has been done in Ref.~\onlinecite{esler}. 
At this stage of development 
the construction of pseudopotentials is 
quite unsatisfactory for high accuracy QMC calculations, 
 since the pseudopotentials  are usually determined 
with  different and less accurate 
techniques, as Hartree-Fock or LDA.  
Despite the recent progress in the use of pseudopotentials within DMC\cite{lrdmc,lrdmc3,lrdmc2},
the implementation of the pseudopotential approximation
in the many-body QMC framework
is not as mature as in the DFT, where 
a remarkable progress was made only after several years of experience with  the so called
 PAW method\cite{blochl,kresse}.
Thus, in QMC we believe there is room 
for a significant improvement to be realized in the next years.

\appendix

\section{Gaussian periodic basis set}
\label{gaussian_periodic}

We use a localized Gaussian basis set on a box of  of lengths $L_x,L_y,L_z$, defining  the periodic electron-ion distance as
\begin{widetext}

\begin{equation}
 r_{iJ}=\sqrt{  \left( \frac{L_x}{\pi}   \sin\left( \frac{\pi}{L_x}  ( x_i -X_j)\right)\right)^2 
+\left( \frac{L_y}{\pi}  \sin\left( \frac{\pi}{L_y}  ( y_i -Y_j)\right)\right)^2+
\left( \frac{L_z}{\pi}  \sin\left( \frac{\pi}{L_z}  ( z_i -Z_j)\right)\right)^2 }
\end{equation}

\end{widetext}

where $x_i,y_i,z_i$ indicate the Cartesian components of the electron coordinates $\textbf{r}_i$ for  $i=1,\cdots N$ and 
$X_j,Y_j,Z_j$ the corresponding ion ones $\textbf{R}_j$ with  $j=1,\cdots N_A$.
The angular part of the Gaussian basis can be defined in 
strict analogy with the 
conventional scheme for open systems.
They are obtained by multiplying the overall Gaussian $ \exp ( -Z r_{iJ}^2 ) $ 
by appropriate polynomials 
 of $  \sin ( \frac{\pi}{L_\mu}  \textbf{r}^\mu_{iJ}) $
and  $ \cos ( \frac{\pi}{L_\mu}  \textbf{r}^\mu_{iJ}) $ , where 
$\textbf{r}^\mu_{iJ} = \textbf{r}^\mu_i - \textbf{R}_j^\mu$ and $\mu=x,y,z$ labels the three Cartesian components.
Strictly speaking in the periodic case there is an arbitrariness in defining this 
polynomials because the multiplication of a polynomial by any even cos-power  
defines an allowed element of the atomic basis satisfying the periodic or 
antiperiodic boundary conditions in each  direction.

In order to define the basis in an accurate and convenient 
way, we have chosen the unique polynomials that contain the minimum
possible cosine powers.
For instance the angular part $Y_{l=2,m} r^2$ of the $d$-wave  orbital is defined by:
\begin{eqnarray}
\textrm{for} &&  m  =  0  : \\
&& 
{\textstyle
\frac{1}{2} \left( 3 \left(\frac{L_3}{\pi} \sin (\frac{\pi}{ L_3}  \textbf{r}^3 )\right)^2 -
r^2\right)
}  \nonumber \\
\textrm{for} && m =  \pm 1 :  \\
&& 
{\textstyle
\sqrt{3} \frac{L_2}{\pi}\sin ( \frac{\pi}{ L_2}  \textbf{r}^2)   
\cos ( \frac{\pi}{ L_2}  \textbf{r}^2) \frac{L_3}{\pi}
\sin (\frac{\pi}{L_3}  \textbf{r}^3)  \cos
( \frac{\pi}{ L_3}  \textbf{r}^3)  
}\nonumber \\
&& 
{\textstyle
\sqrt{3} \frac{L_1}{\pi} \sin  ( \frac{\pi}{ L_1}  \textbf{r}^1)   \cos
( \frac{\pi}{L_1}  \textbf{r}^1) \frac{L_3}{\pi} \sin  ( \frac{\pi}{
    L_3}  \textbf{r}^3) \cos ( \frac{\pi}{ L_3}
  \textbf{r}^3) 
}\nonumber \\
\textrm{for} && m  =  \pm 2 :    \\
&& 
{\textstyle
\sqrt{3} \frac{L_1}{\pi} \sin ( \frac{\pi}{L_1}  \textbf{r}^1)
\cos  ( \frac{\pi}{L_1}  \textbf{r}^1) \frac{L_2}{\pi} \sin  (
  \frac{\pi}{ L_2}  \textbf{r}^2) \cos
( \frac{\pi}{L_2}  \textbf{r}^2) 
} \nonumber \\
 && 
{\textstyle
\frac{\sqrt{3}}{2} \left( \left( \frac{L_1}{\pi} \sin
       (\frac{\pi}{ L_1}  \textbf{r}^1) \right)^2 -
\left( \frac{L_2}{\pi} \sin ( \frac{\pi}{ L_2}
    \textbf{r}^2)\right)^2\right) }\nonumber 
\end{eqnarray}

\begin{acknowledgments}
We acknowledge F. Gygi for his support in using the {\it Qbox} code. 
We thank Shiwei Zhang and Matteo Calandra  for useful discussions and correspondence. 
We are especially in debt with Pasquale Pavone for useful comments and 
suggestions about the phonon calculations. MC thanks the Ecole
Polytechnique, the japanese-french JST-CREST grant for financial support,
and the NCSA computing facilities at the University of Illinois at Urbana-Champaign. 
SS was supported by MIUR COFIN07 and CINECA. 
\end{acknowledgments}

\bibliography{silicon}

\end{document}